\documentclass[12pt,a4paper]{article}
\usepackage[usenames]{color}

\usepackage{amssymb}
\usepackage{amsmath}

\begin{document}
\pagenumbering{arabic}

\def\spade{\par\noindent$\spadesuit$\par\noindent}
\def\club{\par\noindent$\clubsuit$\par\noindent}
\def\heart{\par\noindent$\heartsuit$\par\noindent}
\def\G{\Gamma}

\rightline{IFUM-FT/883}

\noindent
\Large
\bf
\begin{center}
Further Comments on \\
the Symmetric Subtraction of the Nonlinear Sigma Model
\end{center}
\large
\rm
\vskip 1.3 truecm
\centerline{D.~Bettinelli
\footnote{e-mail: {\tt daniele.bettinelli@mi.infn.it}}, 
R.~Ferrari\footnote{e-mail: {\tt ruggero.ferrari@mi.infn.it}}, 
A.~Quadri\footnote{e-mail: {\tt andrea.quadri@mi.infn.it}}}

\normalsize
\medskip
\begin{center}
Dip. di Fisica, Universit\`a degli Studi di Milano\\
and INFN, Sez. di Milano\\
via Celoria 16, I-20133 Milano, Italy
\end{center}

\vskip 0.8 truecm
\bf
\centerline{Abstract}

\normalsize
\rm

\vskip 0.5 truecm
\begin{quotation}
Recently a perturbative theory has been constructed, starting from
the Feynman rules of the  nonlinear sigma model at the tree level 
in the presence of an external vector source
coupled to the flat connection and of a scalar source
coupled to the nonlinear sigma model constraint
(flat connection formalism).

The construction is based on a local functional equation, which
overcomes the problems due to the presence 
(already at one loop) of 
non chiral symmetric divergences. The subtraction procedure 
of the divergences in the loop expansion is
performed by means of minimal subtraction
of properly normalized amplitudes
in dimensional regularization.

In this paper we complete the study of this
subtraction procedure by giving the formal proof
that it is symmetric to all orders
in the loopwise expansion.
 We provide further arguments on the issue that, 
within our subtraction strategy, only
two parameters can be consistently used as physical constants.

\end{quotation}

\newpage
\section{Introduction}
\label{sec:int}
 A quantum field theory based on the Feynman rules
of the nonlinear sigma model is plagued not only by the
presence of an infinite number of superficially divergent amplitudes
but also by the fact that the divergences are not chiral
invariant. These difficulties are present already at the one loop
level,
as has been widely discussed in the existing literature
\cite{Charap:1970xj}-\cite{Appelquist:1980ae}.
Recently the construction of a perturbative theory for
the nonlinear sigma model has been proposed by using a local functional
equation for the generating functional of the 1 PI amplitudes 
\cite{Ferrari:2005ii}.
The equation stems from the invariance under local chiral
transformations of the Haar measure in the path integral.
This formulation overcomes the difficulty due to the lack of
chiral symmetry of the divergences. The subtraction of the
divergences is performed in dimensional regularization
by using minimal subtraction. In the present work
we discuss this subtraction procedure and 
give the formal proof
that it is symmetric to all orders
in the loopwise expansion.
\par
We use the notion of {\sl ``symmetric
subtracted''} 
theory when the perturbation series: i)
can be made finite by the subtraction of the infinities in a
local fashion and ii) the defining functional equation
is not spoiled by the introduction of counterterms. 
Since the defining functional
equation induces  transformations on the
vertex functional in a natural way
(see Section \ref{sec:sub}) and the counterterms 
must have definite invariance properties under these transformations, we use
the adjective {\sl ``symmetric''} in order to indicate the whole of
these essential properties.
\par\noindent
The construction of the perturbative series starts
from the Feynman rules  of the nonlinear
sigma model. The radiative corrections are regularized
by continuation in the dimensions. The strategy by which
the divergences are removed in the limit $D=4$ makes use
of two important properties of the functional equation,
that are duly discussed in Ref. 
\cite{Ferrari:2005ii}, \cite{Ferrari:2005va} and
\cite{Ferrari:2005fc}: 
i) hierarchy
ii) Weak Power Counting (WPC). As summarized in Section \ref{sec:model}
the functional equation has a rigid hierarchy structure in the
loop expansion: all amplitudes involving the pion fields
(descendant amplitudes) are derived from those involving
only insertions of the flat connection ($F_{a\mu}$) and
the order parameter (the constraint $\phi_0$), the ancestor amplitudes.
The important consequence of this fact comes from the second
property: the WPC. At each order of the loop expansion
the number of divergent  ancestor amplitudes
is finite, since the superficial divergence of a graph is
($N_J$ and $N_K$ are the number of flat connection and 
order parameter insertions)
\begin{eqnarray}
(D-2)n +2- N_J - 2 N_{K_0}.
\label{int.1}
\end{eqnarray}
The proof of this  result is recalled in Appendix \ref{app:A}.
Thus at each loop order all amplitudes are made finite
by a finite number of subtractions. Moreover the 
WPC  remains valid only if one does not introduce terms
of higher dimensions in the tree-level Feynman rules.
These facts suggest our subtraction strategy: if one finds
a way to perform subtractions without introducing free
parameters for higher dimensions counterterms in the tree
level Feynman rules, then one gets a consistent theory
with a finite number of physical parameters. 
The subtraction strategy is suggested by the functional equation itself.
The violation of the equation at $n-$th order, when the counterterms
are introduced up to order $n-1$, has simple dimensional properties when
the scale parameter is varied. Then a simple pole removal (minimal
subtraction) automatically restores the functional equation.
\par\noindent
In previous works 
\cite{Ferrari:2005ii}, \cite{Ferrari:2005va} and
\cite{Ferrari:2005fc}
we have discussed this point by means of some
non trivial examples and  of formal arguments. In this work
we
present the proofs of 
the
necessary steps for its implementation.

Under the assumption
of the validity of the local functional equation and 
of the WPC the tree-level Feynman rules turn out to be unique.
These results depend on  the form
of the parametrization of the group
elements in terms of fields  and consequently 
on the particular form of
the transformation on the fields.
Moreover the issue of the number of independent physical
parameters within this perturbative framework can be
addressed. We confirm that only the vacuum expectation
value of the order parameter and the scale of the radiative correction
enter in the final expression of the subtracted amplitudes.

\par

In the case of the nonlinear sigma model the theory is defined
through the effective action $\Gamma$ which has to obey a nonlinear {\sl local}
functional equation. At the one loop level the counterterms 
$\widehat\Gamma ^{(1)}$
obey 
a linearized form of the same equation.  
These counterterms have a particular feature: they are not
present in the vertex functional $\Gamma^{(0)}$ at the tree level. 
Therefore the question arises whether they can be inserted
back into the tree-level vertex functional. 
The answer is negative. 
Some of them do not obey the 
{\sl nonlinear} defining functional equation. Others 
modify in a substantial way
the unperturbed space of states (by introducing ghost states associated
to kinetic terms in $\Box^2$). Finally there are some that
could be introduced in the vertex functional $\Gamma^{(0)}$ at the
tree level, since they obey
the defining local functional equation, but they would spoil the WPC. 

That is, the tree-level vertex functional is uniquely
fixed by the symmetry and the WPC.

Then one can try to assign free parametrs 
to the counterterms at the one loop level. 
Despite this is mathematically allowed,
we argue that this
strategy is not sustainable from the physical point of
view, since parameters should enter in the zero loop
vertex functional
$\Gamma ^{(0)}$. We stress this fundamental point: the expansion parameters
of the classical action might differ from the physical parameters
of the zero loop vertex functional. The presence of a vacuum state
that induces a reshuffling of the perturbative expansion (spontaneous
symmetry breaking) is one example where such a distinction is essential.
\par
After we have excluded free parameters in association to the counterterms,
the question remains of the number of independent parameters in the symmetric subtraction scheme we are proposing. 
One parameter is present in  $\Gamma ^{(0)}$; for instance, 
the vacuum expectation value of
$\phi_0$. However an extra mass parameter can be introduced in order
to perform dimensional subtraction. 
We argue that this parameter has the very important role
of fixing the scale of the radiative corrections.
One can formulate the model in such a way that the dimensional
subtraction scale appears as a front factor of the whole
set of Feynman rules. The final consequence of this
physical requirement is that our subtraction procedure for 
the nonlinear sigma model
depends on two parameters, e.g. the v.e.v. of the
spontaneous breakdown and the dimensional
subtraction scale. 

\par
The present paper is devoted to a detailed illustration of the
above mentioned facts. 
The finding of a symmetric subtraction scheme for the
nonlinear sigma model which is consistent
to all orders in the loopwise expansion allows us
to investigate explicitly one example of 
a nonrenormalizable theory that
can be consistently subtracted (i.e. symmetrically and locally).
\par
The discussion is illustrated at the one loop level, but the necessary
tools for the extension at higher order are also provided. 
In particular we discuss for any order in the loop expansion 
the equation obeyed by counterterms and the consistency of 
the subtraction procedure.

\section{The Nonlinear Sigma Model}
\label{sec:model}

The $D$-dimensional classical action of the nonlinear sigma model in the
flat connection formalism \cite{Ferrari:2005ii} is
\footnote{Here the external current $\vec J_\mu$ of Ref. \cite{Ferrari:2005ii}
has been rescaled by a factor $-\frac{m_D^2}{4}$ and a harmless $J^2$
has been introduced in the effective action.}
\begin{eqnarray}
\G^{(0)} =  \frac{m_D^2}{8} \int d^Dx \, 
\Big ( F^\mu_a - J^\mu_a \Big )^2 + \int d^Dx \, K_0 \phi_0  
\label{appE:1}
\end{eqnarray}
where $m_{ D}= m^{\frac{D}{2}-1}$.
The flat connection is
\begin{eqnarray}&&
F^\mu= F^\mu_a \frac{1}{2}\tau_a = \frac{i}{g} 
\Omega \partial_\mu\Omega^\dagger
\nonumber\\&&
\Omega = \frac{1}{m_D}(\phi_0+ig \tau_a\phi_a)
\label{appE:1.0}
\end{eqnarray}
where  
$\vec{J}_\mu$ is the background connection and $K_0$ is the source
of the constraint $\phi_0$ of the nonlinear sigma model
\begin{eqnarray}
\phi_0^2 + g^2\phi_j^2 = m_D^2 \, .
\label{appE:2}
\end{eqnarray}
$\G^{(0)}$ obeys a $D$-dimensional
local functional equation associated to the local chiral
transformations induced by left multiplication on $\Omega$
by SU(2) matrices
\begin{eqnarray}
U(\omega)\simeq 1 + \frac{i}{2} g\tau_a \omega_a.
\label{appE:22}
\end{eqnarray}
The following equation is required to be valid
for the effective action on the basis of a path integral
formulation of the model
\begin{eqnarray}
- \partial^\mu \frac{\delta \G}{\delta J^\mu_a} 
   - g \epsilon_{abc} J^\mu_b \frac{\delta \G}{\delta J^\mu_c}
+g\frac{1}{2} \epsilon_{abc}\phi_c \frac{\delta \G}{\delta \phi_b} 
+ g^2\frac{1}{2} \phi_a K_0 
+\frac{1}{2} \frac{\delta \G}{\delta K_0} \frac{\delta \G}{\delta
  \phi_a}
  = 0 \, .
\label{appE:2bis}
\end{eqnarray}
The equation is local (no x-integration). The generating functional
of the Green functions obeys the corresponding equation
\begin{eqnarray}
\Big(\partial^\mu \frac{ \delta }{\delta J_a^\mu}  
+ g\epsilon_{abc}J_b^\mu\frac{ \delta }{\delta J_c^\mu}
+ \frac{1}{2} g\epsilon_{abc}K_b \frac{\delta }{\delta K_c}
- \frac{1}{2} g^2  K_0  \frac{\delta }{\delta K_a} 
+ \frac{1}{2} K_a \frac{\delta }{\delta K_0}
\Big)\textit{Z} =0.
\label{appE:4bis}
\end{eqnarray}
The na{\"\i}ve Feynman rules given implicitly in eq. (\ref{appE:1})
yield amplitudes that solve eqs. (\ref{appE:2bis}) and 
(\ref{appE:4bis}). This property
has been conjectured in Ref. \cite{Ferrari:2005ii} and it is proved
in Appendix \ref{app:B}.
\par

The most general solution to eq.(\ref{appE:2bis}) 
in the loopwise expansion has been
characterized by cohomological methods in \cite{Bettinelli:2007kc}.

The spontaneous breakdown of the global chiral symmetry is fixed
by the boundary condition
\begin{eqnarray}
\left .
 \frac{\delta \G}{\delta K_0}\right |_{\vec\phi=\vec J_\mu=K_0=0}=m_D.
\label{mod.6}
\end{eqnarray}
It will be required that these equations ((\ref{appE:2bis}), 
(\ref{appE:4bis})and (\ref{mod.6}))
remain valid also for the subtracted
amplitudes (symmetric subtraction). 
\par
The non linearity of the equation (\ref{appE:2bis})
is responsible for many peculiar facts. In particular 
by eq.(\ref{mod.6}) $\frac{\delta \G}{\delta K_0}$
is invertible as a formal power series. 
Therefore by using eq.(\ref{appE:2bis}) all amplitudes
involving the $\vec\phi$ fields (descendants) 
can be derived from those of $\vec F_{\mu}$ and
$\phi_0$ (ancestors), 
i.e. the functional derivatives with respect to $\vec J_\mu$ and
$K_0$ (hierarchy).
\par
The tree level amplitudes are fixed by the conditions
\begin{eqnarray}
&&
\frac{\delta^2 \G^{(0)}}{\delta J_a^\mu(x)\delta J_b^\nu(y)}
=\frac{m_D^2}{4}g_{\mu\nu}\delta_{ab} \delta_D(x-y)
\nonumber\\&&
\frac{\delta^2 \G^{(0)}}{\delta K_0(x)\delta K_0(y)}=0
\nonumber\\&&
\frac{\delta^2 \G^{(0)}}{\delta K_0(x)\delta J_b^\nu(y)}=0.
\label{mod.7}
\end{eqnarray}
The dependence of the solution from the parameter $g$ 
is somehow peculiar. 
Given an {\sl unsubtracted} solution $\Gamma[\vec J,K_0,\vec \phi,m_D,g]$ 
of  equation (\ref{appE:2bis})
with the boundary conditions (\ref{mod.6}) and (\ref{mod.7}),
one can  check that 
\begin{eqnarray}
\Gamma[g^{-1}\vec J,g^{-1}K_0,\vec \phi,g~m_D,g]
\label{mod.007}
\end{eqnarray}
obeys the same equations with $g=1$. Thus $g$ can be removed  
by a redefinition of the mass scale parameter $m_D\to g~ m_D$ 
(together with  $\vec J  \to g^{-1}\vec J $ and $K_0  \to 
g^{-1}K_0 $), i.e. for unsubtracted vertex functional one has
\begin{eqnarray}
\Gamma[g^{-1}\vec J,g^{-1}K_0,\vec \phi,g~m_D,g]
=
\Gamma[\vec J,K_0,\vec \phi,m_D,1].
\label{mod.007p}
\end{eqnarray}
However
the situation changes if one wants to define the theory at 
$D=4$. Subtraction of poles is needed and, together with this,
a scale parameter in the definition of the Feynman amplitudes is necessary.
At one loop level the dependence
of the subtracted amplitudes from $\ln m$ 
(in $D=4$) does not allow the complete removal of $g$.
 Thus, at least at the one loop level,
the introduction of $g$ is equivalent to use an extra mass
scale in  the dimensional subtraction and accounts for 
 variants of the minimal subtraction. 
\par
There is another interesting rescaling strategy, i.e. consider 
\begin{eqnarray}
\Gamma[g^{-1}\vec J,K_0,g^{-1}\vec \phi,m_D,g].
\label{mod.008}
\end{eqnarray}
This vertex functional satisfies the eq. (\ref{appE:2bis})
with $g=1$ and eq. (\ref{mod.6}) unchanged. But eq.  (\ref{mod.7})
becomes
\begin{eqnarray}
\frac{\delta^2 \G^{(0)}}{\delta J_a^\mu\delta J_b^\nu}
=\frac{m_D^2}{4g^2}g_{\mu\nu}\delta_{ab}
=\left(\frac{m}{g}\right)^2\frac{m^{(D-4)}}{4}g_{\mu\nu}\delta_{ab}
\label{mod.7008}
\end{eqnarray}
i.e., again, we have a new mass parameter
\begin{eqnarray}
v \equiv \frac{m}{g}
.
\label{mod.7009}
\end{eqnarray}
The discussion on the r\^ole of the parameter $g$ will be resumed and
expanded in Sec. \ref{sec:prop}.
\par
The Feynman rules provided by eq.(\ref{appE:1}) give rise to a perturbative
expansion governed by the WPC theorem \cite{Ferrari:2005va}. The superficial
degree of divergence of a $n$-th loop amplitude involving $N_J$
insertions of the flat connection and $N_{K_0}$ insertions of the
nonlinear sigma model constraint is (see Appendix \ref{app:A})
\begin{eqnarray}
\delta = (D-2) n + 2 -  N_J -2 N_{K_0} \, .
\label{wpcnew}
\end{eqnarray}
%
\section{Subtractions at $D=4$. One Loop}
\label{sec:sub}
Subtractions at $D=4$ are performed in dimensional regularization.
At one loop level the counterterms (obtainable from the pole parts of
the vertex functional $\G^{(1)}$) obey a linearized form
of the local equation eq.(\ref{appE:2bis})
\begin{eqnarray}
&& {\cal S}_a(\widehat\G^{(1)})=\Big [
\frac{1}{2} \frac{\delta \G^{(0)}}{\delta \phi_a}
 \frac{\delta }{\delta K_0}
+
\frac{1}{2} \frac{\delta \G^{(0)}}{\delta K_0} \frac{\delta }{\delta \phi_a}
+g\frac{1}{2} \epsilon_{abc}\phi_c \frac{\delta }{\delta \phi_b} 
\nonumber \\
&& - \partial^\mu \frac{\delta }{\delta J^\mu_a} 
   -  g\epsilon_{abc} J^\mu_b \frac{\delta }{\delta J^\mu_c}
\Big ]\widehat\G^{(1)} = 0 \, .
\label{appE:2tris}
\end{eqnarray}
It is easy to trace in eq. (\ref{appE:2tris}) the transformations induced
through the dependence on $\vec J_\mu$ and $\vec\phi$. Further properties can be derived by introducing
the Grassmann parameter $\omega_a$ and the nilpotent operator 
\cite{Ferrari:2005va}
\begin{eqnarray}
&&
s\equiv \int d^Dx \,  \omega_a {\cal S}_a
=
\int d^Dx \, \Big [ \Big ( \frac{1}{2} \omega_a \phi_0
+ \frac{g}{2} \epsilon_{ajk} \phi_j \omega_k \Big )
\frac{\delta }{\delta \phi_a} 
\nonumber \\
&& 
+ \frac{1}{2} \omega_a \frac{\delta \G^{(0)}}{\delta \phi_a} 
\frac{\delta }{\delta K_0} 
+ \Big ( \partial^\mu \omega_a
+ g\epsilon_{aij} J^\mu_i \omega_j \Big ) 
\frac{\delta }{\delta J^\mu_a} \Big ]  
\label{ext.2}
\end{eqnarray}
with
\begin{eqnarray}
s~  \omega_a = - 
\frac{g}{2} \epsilon_{ajk}\omega_j \omega_k  .
\label{ext.2.1}
\end{eqnarray}

We consider the Legendre transform
\begin{eqnarray}
K_a \equiv -\frac{\delta}{\delta \phi_a}\Gamma^{(0)}
=-\frac{\delta}{\delta \phi_a} S_0 + g^2\frac{K_0}{\phi_0}\phi_a
\label{ext.1}
\end{eqnarray}
where $S_0 = \left . \G^{(0)} \right |_{K_0=0}$.
One gets
\begin{eqnarray}
s ~K_{a'} 
=-\frac{g}{2} \epsilon_{a'ak}  \omega_k K_a 
+ \frac{g^2}{2} \omega_{a'} K_0
\label{ext.3}
\end{eqnarray}
and
\begin{eqnarray}
s~K_0 =-\frac{1}{2} \omega_a K_a.
\label{ext.4}
\end{eqnarray}
Then
\begin{eqnarray}&&
\overline K_0 \equiv
K_0\phi_0 +  K_a\phi_a
\nonumber\\&&
= K_0\phi_0 -\phi_a\frac{\delta}{\delta \phi_a} S_0 +g^2
\frac{K_0}{\phi_0}\phi_a\phi_a
\nonumber\\&&
=\frac{m_D^2K_0}{\phi_0}-\phi_a\frac{\delta}{\delta \phi_a} S_0
\label{ext.5}
\end{eqnarray}
is invariant under $s$:
\begin{eqnarray}
s~\overline K_0 =0.
\label{ext.6}
\end{eqnarray}
In terms of the background connection $J_{a\mu}$ and of the
flat connection
\begin{eqnarray}
F^\mu_a = \frac{2}{m_D^{2}}\big(\phi_0 \partial^\mu \phi_a - 
\partial^\mu \phi_0 \phi_a+ g\epsilon_{abc} \partial^\mu \phi_b \phi_c\big) \, 
\label{appE:3}
\end{eqnarray}
the invariant solutions of the linearized functional equation 
which enter at the one loop level read \cite{Ferrari:2005va}

\begin{eqnarray}
&& {\cal I}_1 = \int d^Dx \, \Big [ D_\mu ( F -J )_\nu \Big ]_a \Big [ D^\mu ( F -J )^\nu \Big ]_a  \, , 
\nonumber \\
&& {\cal I}_2 = \int d^Dx \, \Big [ D_\mu ( F -J )^\mu \Big ]_a \Big [ D_\nu ( F -J )^\nu \Big ]_a  \, , 
\nonumber \\
&& {\cal I}_3 = \int d^Dx \, \epsilon_{abc} \Big [ D_\mu ( F -J )_\nu \Big ]_a \Big ( F^\mu_b -J^\mu_b \Big ) \Big ( F^\nu_c -J^\nu_c \Big ) \, ,  \nonumber \\
&& {\cal I}_4 = \int d^Dx \, \Big ( \frac{m_D^2 K_0}{\phi_0} - \phi_a \frac{\delta S_0}{\delta \phi_a} \Big )^2 \, , \nonumber \\
&& {\cal I}_5 = \int d^Dx \, \Big ( \frac{m_D^2 K_0}{\phi_0} - \phi_a \frac{\delta S_0}{\delta \phi_a} \Big ) \Big ( F^\mu_b -J^\mu_b \Big )^2 \, , 
\nonumber \\
&& {\cal I}_6 = \int d^Dx \, \Big ( F^\mu_a -J^\mu_a\Big  )^2
 \Big ( F^\nu_b -J^\nu_b \Big )^2 \, , \nonumber \\
&& {\cal I}_7 = \int d^Dx \, \Big ( F^\mu_a -J^\mu_a\Big  )
   \Big ( F^\nu_a -J^\nu_a\Big  ) 
   \Big ( F_{b\mu} -J_{b\mu} \Big  )
   \Big ( F_{b\nu} -J_{b\nu} \Big  ) \, ,
\label{appE:4}
\end{eqnarray}
where $D_\mu$ denotes the covariant derivative w.r.t $F_{a\mu}$:
\begin{eqnarray}
D_{ab\mu} = \delta_{ab}\partial_\mu  + g\epsilon_{acb} F_{c\mu } \, .
\label{appE:5}
\end{eqnarray}

By dimensional arguments one expects that at one loop the
counterterms (the $1/(D-4)$ pole parts) 
are linear combinations of ${\cal I}_1\dots {\cal I}_7$.
In Ref. \cite{Ferrari:2005va} the linear combination is explicitly
evaluated. On these grounds other solutions of eq.(\ref{appE:2tris})
are excluded, e.g. 
\begin{eqnarray}
\int d^Dx \overline K_0.
\label{mod.72}
\end{eqnarray}

\section{Subtraction at $D=4$. Higher Loops}
\label{sec:higher}
We now discuss the subtraction procedure at higher
loops. The content of this Section has been the subject of conjectures and 
explicit examples in References 
\cite{Ferrari:2005ii}, \cite{Ferrari:2005va} and
\cite{Ferrari:2005fc}.
Here we try to present an organic formulation. The proofs are given in Appendix
\ref{app:B}, \ref{app:C} and \ref{app:D}.
\par
At higher loops the counterterms
obey a more complex equation, since the lower order terms contribute
to a non homogeneous term
\begin{eqnarray}
&& {\cal S}_a(\widehat\G^{(n)})=\Big [
\frac{1}{2} \frac{\delta \G^{(0)}}{\delta \phi_a}
 \frac{\delta }{\delta K_0}
+
\frac{1}{2} \frac{\delta \G^{(0)}}{\delta K_0} \frac{\delta }{\delta \phi_a}
+g\frac{1}{2} \epsilon_{abc}\phi_c \frac{\delta }{\delta \phi_b} 
\nonumber \\
&& - \partial^\mu \frac{\delta }{\delta J^\mu_a} 
   -  g\epsilon_{abc} J^\mu_b \frac{\delta }{\delta J^\mu_c}
\Big ]\widehat\G^{(n)} =  -\frac{1}{2} \sum_{j=1}^{n-1}
\frac{\delta \widehat\G^{(j)}}{\delta K_0} 
\frac{\delta \widehat\G^{(n-j)}}{\delta \phi_a} \,\, .
\label{mod.8}
\end{eqnarray}
The above equation is valid provided that the subtractions are
performed in such a way that
eqs. (\ref{appE:2bis}) and (\ref{appE:4bis}) are preserved. 
By standard arguments (\cite{Gomis:1994he}, \cite{Weinberg}) one can show the
validity of the consistency condition
\begin{eqnarray}
s~ \int d^Dx \, \omega_a\left(
\sum_{j=1}^{n-1}
\frac{\delta \widehat\G^{(j)}}{\delta K_0} 
\frac{\delta \widehat\G^{(n-j)}}{\delta \phi_a}
\right) = 0 \, 
\label{mod.8.1}
\end{eqnarray}
under the assumption that eq.(\ref{mod.8}) is recursively
fulfilled up to order $n-1$.
For the discussion presented in the next Sections it is worth
to outline the arguments that lead to eq. (\ref{mod.8}).
Consider the formal perturbative expansion of the generating functional
of the Feynman amplitudes, where the counterterms 
$\widehat\Gamma^{(j)}$ have been introduced
\begin{eqnarray}&&
\textit{Z}[\vec J,K_0,\vec K] =\exp i\Big ( \Gamma_{\rm INT}^{(0)}
+\sum_{j\ge 1}\widehat\Gamma^{(j)}
\Big)
\Big|_{\phi_a=-i\frac{\delta}{\delta K_a}}
\nonumber \\&&
\exp \frac{1}{2}\int d^Dx~ d^Dy~ K_b(x)D_F(x-y)K_b(y).
\label{mod.10}
\end{eqnarray}
We introduce a shorthand notation 
\begin{eqnarray}
\widehat\G\equiv \Gamma^{(0)}
+\sum_{j\geq 1}\widehat\Gamma^{(j)}.
\label{mod.12}
\end{eqnarray}
In the Appendix \ref{app:B} we give a diagrammatic proof
of the following relation, which essentially shows the
validity of the Quantum Action Principle
{\em \`a la} Breitenlohner and Maison \cite{QAP}
\begin{eqnarray}&&
\Big(-\partial^\mu \frac{ \delta }{\delta J_a^\mu}  
- g\epsilon_{abc}J_b^\mu\frac{ \delta }{\delta J_c^\mu}
+\frac{1}{2} g^2  K_0  \frac{\delta }{\delta K_a} 
 -\frac{1}{2} K_a \frac{\delta }{\delta K_0}
- \frac{1}{2} g\epsilon_{abc}K_b \frac{\delta }{\delta K_c}
\Big)\textit{Z} =
\nonumber \\&&  
i \biggl(-\partial^\mu \frac{\delta 
\widehat\Gamma}{\delta J^\mu_a} 
   - g \epsilon_{abc} J^\mu_b \frac{\delta \widehat\G}{\delta J^\mu_c}
-\frac{1}{2} g^2 K_0 \phi_a 
+\frac{1}{2}\frac{\delta \widehat\G}{\delta K_0} 
\frac{\delta \widehat\G}{\delta \phi_a}
+g\frac{1}{2} \epsilon_{abc}\phi_c \frac{\delta \widehat\G}{\delta \phi_b} 
\biggr)\cdot \textit{Z}\, ,
\nonumber \\&& 
\label{mod.11}
\end{eqnarray}
where the dot indicates the insertion of the local operators.
Eq. (\ref{mod.11}) shows the connection between eqs. (\ref{appE:2bis}) 
and (\ref{appE:4bis}) and eq. (\ref{mod.8}). Equation (\ref{mod.11}) 
for generic  $D$ is valid also without counterterms. 
In this case it shows that the amplitudes constructed with the na{\"\i}ve
Feynman rules generated from $\Gamma^{(0)}$ are solutions of 
eqs. (\ref{appE:2bis}) and (\ref{appE:4bis}) \cite{Ferrari:2005ii}.
\par
Before we describe the subtraction procedure, it is worth to 
illustrate a further
equation for the 1PI generating functionals $\Gamma^{(n)}$. 
We sort the 1PI functionals
according to the
total power in $\hbar$ of the counterterms present in the Feynman integrals. 
Let us then define by
$\Gamma^{(n,k)}$ the n-loop 1PI functional where the power of $\hbar$
of the counterterms
is a fixed $k\leq n$. Then the vertex function at n-loop is
\begin{eqnarray}
\Gamma^{(n)}=\sum_{j=0}^n\Gamma^{(n,j)}.
\label{mod.13}
\end{eqnarray}
In Appendix \ref{app:C} we prove the following equation($n > 0$)
\begin{eqnarray}
&&
\Biggl(
- \partial^\mu \frac{\delta }{\delta J^\mu_a} 
   - g \epsilon_{abc} J^\mu_b \frac{\delta }{\delta J^\mu_c}
+g\frac{1}{2} \epsilon_{abc}\phi_c \frac{\delta }{\delta \phi_b} 
+\frac{1}{2}\frac{\delta \G^{(0)}}{\delta K_0} 
\frac{\delta }{\delta\phi_a}
+
\frac{1}{2}
\frac{\delta \G^{(0)}}{\delta\phi_a}
\frac{\delta }{\delta K_0} 
\Biggr)\G^{(n,k)}
\nonumber \\&& 
+\frac{1}{2} \sum_{n'=1}^{n-1} \, \, 
\sum_{j\ge {\rm max}(0,k-n+n')}^{j\leq {\rm min}(k,n')}
\frac{\delta \G^{(n',j)}}{\delta K_0} 
\frac{\delta \G^{(n-n',k-j)}}{\delta\phi_a}
  = 0 \, .
\nonumber \\&& 
\label{mod.14}
\end{eqnarray}
The consistency condition 
\begin{eqnarray}
&&
s~ \int d^D x \omega_a(x) \left(
 \sum_{n'=1}^{n-1} \, \, 
\sum_{j\ge {\rm max}(0,k-n+n')}^{j\leq {\rm min}(k,n')}
\frac{\delta \G^{(n',j)}}{\delta K_0} 
\frac{\delta \G^{(n-n',k-j)}}{\delta\phi_a}
\right)
  = 0 
\nonumber \\&& 
\label{mod.15}
\end{eqnarray}
is valid also in this case, but the proof will be omitted.
Eq. (\ref{mod.14}) is a very powerful tool for investigations over the
validity of the functional equation (\ref{appE:2bis}), since it allows the
study of the counterterms by introducing a grading on them.
Consider the  1PI generating functional  where counterterms 
$\widehat\Gamma^{(j)}$ have been introduced up to $n-1$ loops
in such a way to fulfill the local functional equation
 and to remove the poles up to order $n-1$.
Then at $n$-loops poles in $D-4$ are present and moreover we
expect a violation of eq.~(\ref{appE:2bis}). In Appendix \ref{app:D}
we show that the breaking is given by the following equation
\begin{eqnarray}
&&  - \partial^\mu \frac{\delta \G^{(n)}}{\delta J^\mu_a} 
   - g \epsilon_{abc} J^\mu_b \frac{\delta \G^{(n)}}{\delta J^\mu_c}
+g\frac{1}{2} \epsilon_{abc}\phi_c \frac{\delta \G^{(n)}}{\delta \phi_b} 
 \nonumber \\&&
+\frac{1}{2}\frac{\delta \G^{(0)}}{\delta K_0} 
\frac{\delta \G^{(n)}}{\delta \phi_a}
+\frac{1}{2}\frac{\delta \G^{(0)}}{\delta \phi_a}
\frac{\delta \G^{(n)}}{\delta K_0} 
+\frac{1}{2}\sum_{j=1}^{n-1} \frac{\delta \G^{(n-j)}}{\delta K_0} 
\frac{\delta \G^{(j)}}{\delta \phi_a}
 \nonumber \\&&
= 
 \frac{1}{2} \sum_{j=1}^{n-1}
\frac{\delta \widehat\G^{(j)}}{\delta K_0} 
\frac{\delta \widehat\G^{(n-j)}}{\delta \phi_a}
\label{mod.9}
\end{eqnarray}
Since the bilinear terms have no poles in 
$D-4$, the procedure of minimal subtraction yields  $n$-loop
counterterms that obey a non homogeneous linearized equation.
These n-th order counterterms obey then eq. (\ref{mod.8}).
\par
Our strategy of subtraction of infinities is based on eq. (\ref{mod.9}).
If we properly normalize the amplitudes, the breaking term in eq.(\ref{mod.9})
contains only poles in $D-4$ (no finite parts!). 
Thus minimal subtraction for the properly
normalized amplitudes removes the breaking term and therefore yields
a recursive and consistent procedure based only on the parameters $v$ 
and $m$, i.e. on the vev and the scale of dimensional regularization.
This subtraction procedure is presented in full details in Appendix
\ref{app:D}.

%
\section{Parameters Fixing}
\label{sec:fixing}
In this Section we show that we cannot introduce at the
tree level new Feynman 
vertices associated to the one-loop counterterms if we want
to produce a sensible and consistent theory.
\par
Minimal subtraction is of course a very interesting option
in order to make finite the perturbative series. The conjecture
that this subtraction algorithm is symmetric 
(i.e. eq. (\ref{appE:2bis}) is stable)
is supported by some general arguments (given in Sec.
\ref{sec:higher}) and by an explicit example
in Ref. \cite{Ferrari:2005fc}. Appendix \ref{app:D} gives
the final proof that the conjecture is indeed correct.
Thus this theory can be tested by experiments.

\par
A frequent objection to the present proposal of making finite a
nonrenormalizable theory is that one needs seven parameter-fixing
appropriate measures
in order to evaluate the coefficients of 
 ${\cal I}_1\dots {\cal I}_7$. This objection is legitimate if
the above mentioned invariants are action-like. As  one should do
in power counting
renormalizable theories, according to algebraic
renormalization \cite{Symanzik:1969ek}-\cite{fgq}. Here the situation is more involved. This is
evident if we paraphrase the problem in the following
way. Can we introduce at the tree level the seven
invariants with arbitrary coefficients and treat them as
{\sl bona fide} interaction terms intervening in the loop
expansion as the original one provided in $\G^{(0)}$ of
eq. (\ref{appE:1})?  The answer to this question is in general
negative. If one allows this modification of the unperturbed effective
action, the one loop corrections will be modified by extra
terms generated by the newly introduced Feynman rules, thus
bringing to a never ending story.
\par
In particular the introduction at tree level of the vertices
described by the invariants in eq.(\ref{appE:4}) implies
new Feynman rules which invalidate the weak power-counting
\cite{Ferrari:2005va} 
(with the exception of the 
combination
\begin{eqnarray}
2 ({\cal I}_1 - {\cal I}_2) - 4 {\cal I}_3 + {\cal I}_6 - {\cal I}_7 =
\int d^Dx \, G_{a\mu\nu} G_a^{\mu\nu} \, ,
\label{comm.add}
\end{eqnarray}
which however depends only on the field strength squared 
of 
the external source $J_{a\mu}$ 
and thus does not modify the Feynman rules
for the pions).

The superficial degree of divergence of the ancestor
amplitudes is not any more given by eq. (\ref{wpcnew}).
As a direct consequence of the violation of the weak power-counting,
already at one loop the number of divergent ancestor amplitudes
is infinite.
\par
A closer look to ${\cal I}_1\dots {\cal I}_7$ shows that 
there are also other reasons that forbid the use of some of these invariants
as unperturbed effective action terms.
${\cal I}_1,{\cal I}_2$ can be introduced into
$\G^{(0)}$ without breaking eq. (\ref{appE:2bis}). However
unless they appear in the combination ${\cal I}_1-{\cal I}_2$
(see eq.(\ref{comm.add}))
they modify the spectrum of the unperturbed states (by introducing
negative norm states) through kinetic terms with four derivatives.
${\cal I}_4, {\cal I}_5$ cannot be introduced into $\G^{(0)}$
because they violate eq. (\ref{appE:2bis}).

\section{Finite Subtractions}
\label{sec:fin}
After we excluded the possibility of introducing
in the tree level effective action the invariants ${\cal I}_1\dots {\cal I}_7$,
there is still the possibility to use them for a finite, in principle
arbitrary, renormalization
{\em strictly at one loop level}. I.e. in the book keeping of
the Feynman rules one could enter new terms
\begin{eqnarray}
\hbar~\sum_j \lambda_j \int d^D x~{\cal I}_j(x) \, ,
\label{fin.1}
\end{eqnarray}
where we have explicitly exhibited the $\hbar$ factor in order
to remind that these vertices are of first order in $\hbar$
expansion. $\lambda_j$ are arbitrary real parameters. 
\par
More explicitly we can tell the story in the following way. The
subtraction of the poles in $D-4$ requires a series of counterterms
of the form (\ref{fin.1}) where the coefficients carry the
pole factor $1/(D-4)$. Then the option  to use these
extra degrees of freedom as free parameters can be explored
since it is mathematically allowed. 

In the case 
of a power counting renormalizabel theory
the fixing of the finite parts of
the symmetric counterterms can be seen as a way
to introduce the renormalization by a reset of the 
parameters entering into the classical action. 
The situation is clearly different in the present case, 
since the invariants ${\cal I}_1,\dots,
{\cal I}_7$ are not action-like and therefore
the additional parameters $\lambda_j$ can be introduced 
only as independent quantum corrections.

The meaning of this latter procedure 
seems to us rather unclear 
from the physical point of view.
If one wishes to introduce new independent parameters
one should do so at tree level (notice that
the local functional equation is non-anomalous). 
If one requires in addition the WPC, then
there is no more freedom left and one ends
up with the tree-level Feynman rules
encoded in eq.(\ref{appE:1}).

\section{Normalization of the amplitudes}
\label{sec:prop}

In minimal subtraction we use pure pole subtraction 
in order to make the theory finite in $D=4$.
Even with this clear cut strategy, still there is some
freedom left connected to the presence of $g$ or
equivalently to the use of a second scale parameter in 
the Feynman rules in dimensional renormalization.
Here we would like to give a formulation of this choice
that has some appeal.
\par
When we evaluate the counterterms, by starting
from the vertex functional in $D$ dimensions, we automatically
make a statement  on their finite parts. Thus from a generic
amplitude in $D$ dimensions involving $n$ external currents $J$
\begin{eqnarray}
\Gamma[J_1\cdots J_n|D]
\label{fin.2}
\end{eqnarray}
the counterterm is obtained by using the normalized function
\begin{eqnarray}
\frac{m^2}{m_D^{2}}\Gamma[J_1\cdots J_n|D]=
\frac{1}{m^{(D-4)}}\Gamma[J_1\cdots J_n|D].
\label{fin.3}
\end{eqnarray}
Its pole part in $D=4$ fixes the counterterms. For example, the
single pole part in $\Gamma[J_1\cdots J_n,|D]$ is removed by the
counterterm mechanism
\begin{eqnarray}
\Gamma[J_1\cdots J_n,|D]-  \frac{m^{(D-4)}}{(D-4)}~\lim_{D'\to 4}
\left((D'-4)m^{-(D'-4)}
\Gamma[J_1\cdots J_n,|D']\right).
\label{fin.4}
\end{eqnarray}
The normalization used in eq. (\ref{fin.3}) is needed in order
to produce the correct dimensions of the counterterms in eq.
(\ref{fin.4}). Similarly one proceeds with $K_0$. The normalized
function is
\begin{eqnarray}
\frac{m^{D-4}}{m^{n(\frac{D}{2}-2)}}\Gamma[K_{01}\cdots K_{0n}|D]
=m^{(1-\frac{n}{2})(D-4)}\Gamma[K_{01}\cdots K_{0n}|D].
\label{fin.3.1}
\end{eqnarray}
\par
Eq. (\ref{fin.3}) show that the parameter $g$ can be removed
only in unsubtracted amplitudes (i.e. at $D\not =4$). 
At the one loop level the replacement
$m_D \to g~m_D$ ($m\to m g^\frac{2}{D-2}$)leads to a $\ln g$ 
dependence of the amplitude in $D=4$. In fact eq. (\ref{fin.3})
becomes
\begin{eqnarray}
\left(m\, g^\frac{2}{D-2}\right)^{(4-D)}\Gamma[J_1\cdots J_n|D].
\label{fin.3.2}
\end{eqnarray}
Thus the minimal subtraction introduces in this case
a new mass scale  $ mg$.
\par
The formulation with two parameters takes a particularly elegant form
if we suppress $g$ in favor of a second mass scale 
and moreover
we assign to $K_0$ a dimension that is $D$-independent; i.e. in a way
that the normalization factor for the subtraction of the
poles is identical both for  $\vec J_\mu$ and $K_0$.
To achieve
this normalization we perform a  transformation similar to eq. (\ref{mod.008})
\begin{eqnarray}
\Gamma[g^{-1}\vec J,\left(\frac{gm}{m_D}\right)^{-1}K_0,
\left(\frac{gm}{m_D}\right)g^{-1}\vec\phi,m_D,g].
\label{mod.008.1p}
\end{eqnarray}
\par\noindent
Thus one gets
\footnote{
Note that $v$ cannot be removed by a rescaling $\vec\phi\to
v~\vec\phi$
and $K_0\to 1/v K_0$. In fact the dependence on $v$ remains in eq.
(\ref{mod.7008.1p}).
}
\begin{eqnarray}
- \partial^\mu \frac{\delta \G}{\delta J^\mu_a} 
   -  \epsilon_{abc} J^\mu_b \frac{\delta \G}{\delta J^\mu_c}
+\frac{1}{2} \epsilon_{abc}\phi_c \frac{\delta \G}{\delta \phi_b} 
-\frac{1}{2} \phi_a K_0 
+\frac{1}{2}
 \frac{\delta \G}{\delta K_0} \frac{\delta \G}{\delta
  \phi_a}
  = 0 ,
\label{fin.4.2.1p}
\end{eqnarray}
\begin{eqnarray}
\frac{\delta^2 \G^{(0)}}{\delta J_a^\mu\delta J_b^\nu}
=\frac{v^2}{4}m^{(D-4)}g_{\mu\nu}\delta_{ab}
\label{mod.7008.1p}
\end{eqnarray}
and
\begin{eqnarray}
\left .
 \frac{\delta \G}{\delta K_0}\right |_{\vec\phi=\vec J_\mu=K_0=0}
=\left(\frac{m_D}{gm}\right)~m_D = v m^{(D-4)}.
\label{fin.4.3p}
\end{eqnarray}

This amounts to formally perform the path-integral according to 
(${\cal D} \Omega$ denotes the invariant Haar measure over SU(2))
\begin{eqnarray}
&& 
\!\!\!\!\!\!\!\!\!\!\!\!\!\!\!\!\!\!\!\!\!\!
Z[\vec{J},K_0,\vec{K}] 
\nonumber \\
&&  
\!\!\!\!\!\!\!\!\!\!\!\!\!\!\!\!
=  \int {\cal D}\Omega \exp 
 i ~\int d^Dx \biggl\{ m^{(D-4)}\left[ 
\frac{v^2}{8} \, 
\left ( F^\mu_a - J^\mu_a \right )^2\right] +  K_0 \phi_0  
+ K_a \phi_a
\biggr \} 
\label{fin.4.1}
\end{eqnarray}
with
\begin{eqnarray}&&
F^\mu= F^\mu_a \frac{1}{2}\tau_a = i
\Omega \partial_\mu\Omega^\dagger
\nonumber\\&&
\Omega = \frac{1}{v m^{D-4}}(\phi_0+i \tau_a\phi_a)
\label{fin.4.1.0}
\end{eqnarray}
and
\begin{eqnarray}
\phi_0^2 + \phi_j^2 = v^2 m^{2(D-4)} \, .
\label{fin.4.1.1}
\end{eqnarray}
\par
By this choice the dimensions
of $\vec J_\mu$  and $K_0$ are equal to one and three respectively. The evaluation of
the counterterms is then the same (independently from the
number of $\vec J_\mu$ and of  $K_0$ ), via simple pole subtraction
of the normalized functions as in eq. (\ref{fin.3})
\begin{eqnarray}
\left(\frac{1}{m}\right)^{(D-4)}\Gamma[J_1\cdots J_n
K_{01}\cdots K_{0n'}|D].
\label{fin.4.4}
\end{eqnarray}
Then the full set of Feynman rules is 
\begin{eqnarray}
\widehat\G=\int d^Dx \Biggl\{m^{(D-4)}\Biggl( 
\frac{v^2}{8} \, 
\left ( F^\mu_a - J^\mu_a \right )^2 
+\sum_{j=1}{\cal M}^{(j)}\Biggr)+  K_0 \phi_0  
\Biggr\},
\label{fin.4.5p}
\end{eqnarray}
where the ${\cal M}^{(j)}$  are the local counterterms containing
the pole parts in $D=4$. The (non trivial) finite parts of the subtractions
are governed by the sole front factor $m^{-(D-4)}$ in eq. (\ref{fin.4.4}).
The resulting amplitudes depend on the parameters $v$
and $m$. The last one is not present in the classical action at $D=4$:
it sneaks in as a scale of the radiative corrections.

A similar mechanism has a renowned antecedent in the theory
of Lamb shift \cite{bethe}, where the radiative corrections due to the
excited state transitions  need a ultraviolet cut-off
which is not present at the lowest level of the theory of 
the Hydrogen atom.

A comment is in order here.
In the NLSM a shift in $m$ cannot be 
compensated by a shift in $v$.
Therefore $m$ has to be treated as a second independent 
free parameter (in addition to $v$)
to be determined through the fit with the experimental data.

\par
The question whether this subtraction can be performed by means of other regularization schemes has been considered. Only limited results
have been achieved. One difficulty is related
to the fact that
all these schemes spoil the defining local
functional equation. 
By using the renormalized linear sigma model 
in the limit of large
coupling constant one can get, after subtraction of divergent terms,
the nonlinear sigma model we are proposing (one loop has been checked
in ref. \cite{Bettinelli:2006ps}). This requires a fine tuning in the 
finite subtractions and consequently there is no evidence for a particular
advantageous choice  in the finite subtraction as in dimensional 
regularization.
In order to study this issue it is very useful to consider
the most general solution allowed by the linearized homogeneous 
functional equation. At one loop this means seven
arbitrary coefficients associated to the invariants
reported in Sect.~\ref{sec:sub} (see eq.(\ref{appE:4})).
The same pattern is present in other regularization 
procedures as Pauli-Villars.

\section{Conclusions}
\label{sec:conc}
In this paper it has been proven 
that the subtraction
procedure based on the flat connection
formalism is indeed consistent (i.e. local
and symmetric) to all orders in the loop
expansion.

This subtraction scheme is based on some novel technical tools. First the
use of the vertex functional instead of the action, in order to formulate
the theory. A functional equation is then the defining instrument. 
This equation has
some essential properties: hierarchy and weak power counting. These properties
are the guiding tracks for the construction of  a perturbative expansion
in the number of loops. The restoration of the functional equation at
every oder at $D=4$ suggests a subtraction procedure which defines the theory
itself. The procedure is based on minimal subtraction 
in dimensional regularization.

The resulting theory depends on $m$ and $g$ (or $v$ if one
uses two scales as in Section \ref{sec:prop}). 
We have also shown that in the subtraction procedure
a mass parameter enters as a scale of the
radiative corrections. In Section \ref{sec:prop} 
we formulated the symmetrically
subtracted nonlinear sigma model in such a way that the
second parameter enters as a common front factor of the
whole Feynman rules (counterterms included). 

It is a remarkable fact that the tree-level Feynman rules
are uniquely fixed by the local functional equation
and the weak power-counting.

Uniqueness of the tree-level Feynman rules
has some important consequences.
In fact at each  order
of the perturbative series one can introduce finite renormalizations
by using the appropriate local solutions of eq. (\ref{mod.8}).
Yet these renormalizations
 cannot be inserted back into the tree-level
vertex functional.
Thus they cannot be interpreted as additional
physical parameters in the loopwise expansion.

\section{Acknowledgments}
One of us (R.F.) is very much indebted to A.A. Slavnov for 
stimulating discussions. A.Q. acknowledges useful
discussions with G.Barnich and the warm hospitality
of the Service de Physique Th\'eorique et Math\'ematique
at the Universit\'e Libre de Bruxelles, Belgium.  

\appendix

\section{Weak Power Counting}

\label{app:A}

Consider amplitudes involving only the flat connection
$F_{a\mu}$ and the order parameter field $\phi_0$.
The number of loops $n$ is given by
\begin{eqnarray}
n= I- \sum_j V^J_j - \sum_k V^{K_0}_k - \sum_l V^{\phi}_l +1
\label{appA:1}
\end{eqnarray}
where $I$ is the number of internal $\vec\phi$-lines, $V^J_j$
the number of vertices with one $J$ and $j$ $\vec\phi$, $V^{K_0}_k$
the number of vertices with one $K_0$ and $k$ $\vec\phi$, $V^{\phi}_l$
the number of vertices with $l$ $\vec\phi$.
\par
The superficial degree of divergence is
\begin{eqnarray}&&
\delta = n~ D - 2~I + \sum_j V^J_j+2\sum_l V^{\phi}_l
\nonumber\\&&
=n~ D+ \sum_j V^J_j-2\biggl( n + \sum_j V^J_j+ \sum_k V^{K_0}_k-1\biggr)
\nonumber\\&&
=n(D-2)+2 - N_J - 2~N_{K_0}
\label{appA:2}
\end{eqnarray}
where
\begin{eqnarray}
N_J \equiv \sum_j V^J_j, \qquad 
N_{K_0}\equiv \sum_k V^{K_0}_k.
\label{appA:3}
\end{eqnarray}
It should be remarked that the superficial degree of divergence does
not depend on the number of $\vec\phi$ self-interaction vertices.

\section{Perturbative Solutions of the Functional Equation}

\label{app:B}

In this Appendix we provide a diagrammatic proof of eq. (\ref{mod.11}).
We follow a technique suggested in Ref. \cite{Ferrari:2005ii}, Section 13.
The framework is given by dimensional regularization, where the
Feynman rules are given by a formal series of local operators
as in eq. (\ref{mod.12}). The propagator and the vertices are originated
from the partition of $\widehat\Gamma$ into a free bilinear term 
$\Gamma^{(0)}_{\rm BIL}$ and the
rest $\widehat\Gamma_{\rm INT}$ 
which yields the interaction and the counterterms.
\par
Consider the following operation 
\begin{eqnarray}&&
\Big(-\partial^\mu \frac{ \delta }{\delta J_a^\mu}  
- g\epsilon_{abc}J_b^\mu\frac{ \delta }{\delta J_c^\mu}
+ \frac{1}{2} g^2  K_0  \frac{\delta }{\delta K_a} 
- \frac{1}{2} K_a \frac{\delta }{\delta K_0}
- \frac{1}{2} g\epsilon_{abc}K_b \frac{\delta }{\delta K_c}
\Big)\textit{Z}
\nonumber \\&& 
\label{appB:1}
\end{eqnarray}
on  the generating functional
\begin{eqnarray}
\textit{Z}[\vec J, K_0, \vec K] 
= \int  \textit{D}[\Omega]\exp i\biggl(\widehat\Gamma[\vec J, K_0, \vec \phi]
+\int d^D x K_a\phi_a
\biggr).
\label{appB:2}
\end{eqnarray}
The functional derivatives are local insertions according
to the formalism of path integral
\begin{eqnarray}&&
 \frac{ \delta \textit{Z}}{\delta J_a^\mu(x)}  
= i \frac{\delta 
\widehat\Gamma}{\delta J^\mu_a(x)}\cdot \textit{Z} 
\equiv i\int  \textit{D}[\Omega]
\frac{ \delta \widehat\Gamma}{\delta J_a^\mu(x)}
\exp i\biggl(\widehat\Gamma
+\int d^D y K_a\phi_a
\biggr)
\nonumber \\&& 
\frac{\delta\textit{Z} }{\delta K_a(x)}
= i\phi_a(x) \cdot \textit{Z}
\equiv i\int  \textit{D}[\Omega]\phi_a(x)
\exp i\biggl(\widehat\Gamma
+\int d^D y K_a\phi_a
\biggr).
\label{appB:3}
\end{eqnarray}
Continuation in $D$ dimensions guarantees the validity 
of eqs. (\ref{appB:3}).
Let us consider
\begin{eqnarray}&&\Biggl[
\frac{\delta}{\delta K_{a_1}(x_1)}\cdots
\frac{\delta}{\delta K_{a_n}(x_n)}\Biggl(
\frac{\delta\widehat\Gamma}{\delta \phi_a(x)}\cdot \textit{Z}
\Biggr)\Biggr]\Biggr|_{\vec K=0}
\nonumber \\&& 
=i^n\Biggl(
\frac{\delta\widehat\Gamma}{\delta \phi_a(x)}
\phi_{a_1}(x_1)\cdots\phi_{a_n}(x_n)
\cdot \textit{Z}
\Biggr)\Biggr|_{\vec K=0}
\nonumber \\&& 
=-i^{n-1}\sum_{j=1}^n \delta_{aa_j}\delta(x-x_j)
\Biggl(
\phi_{a_1}(x_1)\cdots{\widehat \phi_{a_j}(x_j)}\cdots\phi_{a_n}(x_n)
\cdot \textit{Z}
\Biggr)\Biggr|_{\vec K=0}
\label{appB:4}
\end{eqnarray}
where the last step is a consequence of the equation of motion
for $\vec\phi$. In fact one finds
\begin{eqnarray}&&
\Biggl(
\frac{\delta\widehat\Gamma}{\delta \phi_a(x)}
\phi_{a_1}(x_1)\cdots\phi_{a_n}(x_n)
\cdot \textit{Z}
\Biggr)\Biggr|_{\vec K=0}
\nonumber \\&& 
= \int d^Dz \frac{\delta^2\Gamma^{(0)}_{\rm BIL}}
{\delta \phi_a(x)\delta \phi_b(z)}\Biggl(\phi_b(z)
\phi_{a_1}(x_1)\cdots\phi_{a_n}(x_n)
\cdot \textit{Z}
\Biggr)\Biggr|_{\vec K=0}
\nonumber \\&& 
+
\Biggl(
\frac{\delta\widehat\Gamma_{\rm INT}}{\delta \phi_a(x)}
\phi_{a_1}(x_1)\cdots\phi_{a_n}(x_n)
\cdot \textit{Z}
\Biggr)\Biggr|_{\vec K=0}.
\label{appB:4.1}
\end{eqnarray}
Finally we perform  the relevant contractions
\begin{eqnarray}&&
= i\sum_{j=1}^n \delta_{aa_j}\delta(x-x_j)
\Biggl(
\phi_{a_1}(x_1)\cdots{\widehat \phi_{a_j}(x_j)}\cdots\phi_{a_n}(x_n)
\cdot \textit{Z}
\Biggr)\Biggr|_{\vec K=0}
\nonumber \\&& 
+
i\int d^Dz \frac{\delta^2\Gamma^{(0)}_{\rm BIL}}
{\delta \phi_a(x)\delta \phi_b(z)}\int d^Dz' \langle
T(\phi_b(z)\phi_{b'}(z'))
\rangle
\nonumber \\&& 
\Biggl(
\frac{\delta\widehat\Gamma_{\rm INT}}{\delta \phi_{b'}(z')}
\phi_{a_1}(x_1)\cdots\phi_{a_n}(x_n)
\cdot \textit{Z}
\Biggr)\Biggr|_{\vec K=0}
\nonumber \\&& 
+
\Biggl(
\frac{\delta\widehat\Gamma_{\rm INT}}{\delta \phi_a(x)}
\phi_{a_1}(x_1)\cdots\phi_{a_n}(x_n)
\cdot \textit{Z}
\Biggr)\Biggr|_{\vec K=0}
\nonumber \\&& 
= i\sum_{j=1}^n \delta_{aa_j}\delta(x-x_j)
\Biggl(
\phi_{a_1}(x_1)\cdots{\widehat \phi_{a_j}(x_j)}\cdots\phi_{a_n}(x_n)
\cdot \textit{Z}
\Biggr)\Biggr|_{\vec K=0}.
\label{appB:4.1.1}
\end{eqnarray}
Eq. (\ref{appB:4}) shows that
\begin{eqnarray}
\frac{\delta\widehat\Gamma}{\delta \phi_a(x)}\cdot \textit{Z}
=-K_a(x)\textit{Z}.
\label{appB:5}
\end{eqnarray}
In order to apply the result of eq. (\ref{appB:5}) to the 
expression in eq. (\ref{appB:1}), one has
to consider the situation where
\begin{eqnarray}
x=x_j.
\label{appB:6}
\end{eqnarray}
These contributions have to be neglected
since the massless propagator for coinciding points is zero
in dimensional regularization
\begin{eqnarray}
D_F(0)= \frac{i}{(2\pi)^D}\int \frac{d^D k}{k^2+i\epsilon}=0.
\label{appB:7}
\end{eqnarray}
Then eq. (\ref{appB:3}) shows together with eq. (\ref{appB:5})
that 
\begin{eqnarray}&&
\Big(-\partial^\mu \frac{ \delta }{\delta J_a^\mu}  
- g\epsilon_{abc}J_b^\mu\frac{ \delta }{\delta J_c^\mu}
+ \frac{1}{2} g^2  K_0  \frac{\delta }{\delta K_a} 
- \frac{1}{2} K_a \frac{\delta }{\delta K_0}
- \frac{1}{2} g\epsilon_{abc}K_b \frac{\delta }{\delta K_c}
\Big)\textit{Z} =
\nonumber \\&&  
i \biggl(-\partial^\mu \frac{\delta 
\widehat\Gamma}{\delta J^\mu_a} 
   - g \epsilon_{abc} J^\mu_b \frac{\delta \widehat\G}{\delta J^\mu_c}
+\frac{1}{2} g^2 K_0 \phi_a  
+\frac{1}{2}\frac{\delta \widehat\G}{\delta K_0} 
\frac{\delta \widehat\G}{\delta \phi_a}
+\frac{1}{2} g\epsilon_{abc}\phi_c \frac{\delta \widehat\G}{\delta \phi_b}
\biggr)\cdot \textit{Z}
\nonumber \\&& 
\label{appB:8}
\end{eqnarray}
i.e. one gets eq. (\ref{mod.11}). 
\par
If  the counterterms obey the equation
\begin{eqnarray}&&
-\partial^\mu \frac{\delta 
\widehat\Gamma}{\delta J^\mu_a} 
   - g \epsilon_{abc} J^\mu_b \frac{\delta \widehat\G}{\delta J^\mu_c}
+\frac{1}{2} g^2 K_0 \phi_a  
+\frac{1}{2}\frac{\delta \widehat\G}{\delta K_0} 
\frac{\delta \widehat\G}{\delta \phi_a}
+\frac{1}{2} g\epsilon_{abc}\phi_c \frac{\delta \widehat\G}{\delta \phi_b}
 =0,
\nonumber \\&& 
\label{appB:9}
\end{eqnarray}
then eq. (\ref{appE:4bis}) is valid order by order. Thus
eq. (\ref{mod.8}) has to be imposed on the counterterms.
\par
It should be stressed that no special
requirements are imposed on $\widehat\Gamma$. In particular
the counterterms might be absent. In this case eq. (\ref{appB:8}) 
proves that the construction of the perturbative series
in $D$ dimension based on the Feynman rules of the
nonlinear sigma model (without subtractions)
yields a solution of the functional equation
(\ref{appE:2bis})\cite{Ferrari:2005ii}. In fact $\Gamma^{(0)}$
obeys eq. (\ref{appB:9}).


\section{Grading by the Counterterms}

\label{app:C}

This Appendix is devoted to the proof of eq. (\ref{mod.14}). In
Appendix \ref{app:B} we proved that the functional equation
(\ref{appE:4bis}) for the Feynman amplitudes is satisfied 
in $D$ dimensions, if the Feynman rules, collectively denoted
by $\widehat\Gamma$ in eq. (\ref{mod.12}), obey the equation
(\ref{mod.8}). The index $j$ in eq. (\ref{mod.12}) denotes the
grading in the $\hbar$ expansion.
\par
One can change the Feynman rules by using a real parameter $\rho$
\begin{eqnarray}
\widehat\G_\rho=\Gamma^{(0)}
+\sum_{j\geq 1}~\rho^j~\widehat\Gamma^{(j)}
\label{appC:1}
\end{eqnarray}
and eq. (\ref{mod.8}) is still valid.
By using  $\widehat\Gamma_\rho$ one gets a new set of Feynman
amplitudes generated by $\textit{Z}[\rho,\vec J,K_0,\vec K]$
which obey eq. (\ref{appE:4bis}). The corresponding 1PI 
$\Gamma[\rho,\vec J,K_0,\vec \phi]$ satisfies eq. (\ref{appE:2bis}).
By construction only $\Gamma[\rho,\vec J,K_0,\vec
\phi]\bigl|_{\rho=1}$ has a finite limit for $D\to 4$.
\par
For practical calculations it is useful to exploit the fact
that $\Gamma[\rho,\vec J,K_0,\vec \phi]$ is a solution
of the functional equation (\ref{appE:2bis}) for any
real $\rho$. As in reference \cite{Ferrari:2005fc} we introduce the
notation
\begin{eqnarray}
\Gamma^{(n)}_\rho = \sum_{j=0}^n\Gamma^{(n,j)}\rho^j,
\label{appC:2}
\end{eqnarray}
where the exponent of $\rho$ counts the total power of $\hbar$
of the counterterms. By inserting this expression
in eq. (\ref{appE:2bis}) one gets (for $n>0$)
\begin{eqnarray}&&
 \sum_{k=0}^n\rho^k\Biggl(
- \partial^\mu \frac{\delta}{\delta J^\mu_a} 
   - g \epsilon_{abc} J^\mu_b \frac{\delta }{\delta J^\mu_c}
+g\frac{1}{2} \epsilon_{abc}\phi_c \frac{\delta }{\delta \phi_b}
\Biggr )\Gamma^{(n,k)}
\nonumber \\&& 
+\frac{1}{2}\sum_{n'=0}^n \sum_{j=0}^{n-n'} \sum_{j'=0}^{n'}\rho^{j+j'}
\frac{\delta \Gamma^{(n-n',j)}}{\delta K_0} 
\frac{\delta \Gamma^{(n',j')}}{\delta
  \phi_a}
  = 0 \, .
\label{appC:3}
\end{eqnarray}
Since $\rho$ is an arbitrary parameter, one gets
\begin{eqnarray}&&
 \Biggl(
- \partial^\mu \frac{\delta}{\delta J^\mu_a} 
   - g \epsilon_{abc} J^\mu_b \frac{\delta }{\delta J^\mu_c}
+g\frac{1}{2} \epsilon_{abc}\phi_c \frac{\delta }{\delta \phi_b}
\Biggr )\Gamma^{(n,k)}
\nonumber \\&& 
+\frac{1}{2}\sum_{n'=0}^n \sum_{j={\rm max}(0,k-n+n')}^{{\rm min}(n',k)}
\frac{\delta \Gamma^{(n',j)}}{\delta K_0} 
\frac{\delta \Gamma^{(n-n',k-j)}}{\delta
  \phi_a}
  = 0 \, ,
\label{appC:4}
\end{eqnarray}
i.e. equation (\ref{mod.14}).


\section{The Subtraction Procedure}

\label{app:D}
In this Appendix we describe in details how the subtraction
of the divergences are performed in order to take the limit
$D\to 4$. The naturalness of the
procedure has induced us to propose it as a rule in the construction
of a physical theory {\sl tout court}.
\par
We use the Feynman rules in eq. (\ref{fin.4.1}), where $g$ has
been traded  by $v$ according to eq. (\ref{mod.7009}).
This choice of parameters has the advantage that we can keep trace
of the dimensions of the counterterms in terms of powers of $m$, the
scale of the radiative corrections.
\par
After rescaling
\begin{eqnarray}
\phi_a \to  m^{(D-4)} \phi_a
\label{appD:1}
\end{eqnarray}
we get the tree level vertex functional
\begin{eqnarray}&& 
\!\!\!\!\!\!\!\!\!\!\!\!\!\!\!\!\!\!\!
 m^{(D-4)}\int d^Dx 
\biggl\{
\frac{1}{2} \, 
\biggl  ( \partial_\mu\phi_0\partial^\mu\phi_0
+\partial_\mu\phi_a\partial^\mu\phi_a - \frac{v^2}{2}F^\mu_aJ_{a\mu}
+\frac{v^2}{4}J^2 \biggr ) +  K_0 \phi_0  
\biggr \}
\nonumber \\&& 
\!\!\!\!\!\!\!\!\!\!\!\!\!\!\!\!\!\!
= m^{(D-4)}\int d^Dx \biggl\{
\frac{1}{2} \, 
\biggl (
\partial_\mu\phi_a\partial^\mu\phi_a 
+ \frac{\phi_a\partial_\mu\phi_a \phi_b\partial^\mu\phi_b }{\phi_0^2}
- \frac{v^2}{2}F^\mu_aJ_{a\mu}
\nonumber \\&& 
+\frac{v^2}{4}J^2 \biggr ) +  K_0 \phi_0  
\biggr \}
\nonumber \\&& 
\label{appD:2}
\end{eqnarray}
with
\begin{eqnarray}&& 
\phi_0=\sqrt{v^2-\vec\phi^2}
\nonumber \\&& 
\Omega = \frac{1}{v}\left(\phi_0+i\tau_a \phi_a\right)
\nonumber \\&& 
F^\mu_a =  \frac{2}{v^2}\big(\phi_0 \partial^\mu \phi_a - 
\partial^\mu \phi_0 \phi_a+ \epsilon_{abc} \partial^\mu \phi_b \phi_c\big).
\label{appD:3}
\end{eqnarray}
Then the $\vec\phi$ propagator has a factor $ m^{-(D-4)}$, while every
vertices $J-\vec\phi^j$, $K_0-\vec\phi^k $ and $\vec\phi^l $ (see
the notations in Appendix \ref{app:A}) has a factor $ m^{(D-4)}$.
\subsection{One Loop}
The one loop 1PI amplitudes have total power of $m$ equal to zero.
Since the counterterms for ancestor amplitudes 
in eq. (\ref{fin.4.5p}) are of the form
\begin{eqnarray}
\widehat\Gamma^{(1)} =  m^{(D-4)}
\int d^D x~ {\cal M}^{(1)}[J, K_0](x)
\label{appD:4}
\end{eqnarray}
the dimensional subtraction has to be performed on the normalized
vertex functional for the ancestor amplitudes
\begin{eqnarray}
\int d^D x {\cal M}^{(1)}[J, K_0](x) = -\frac{1}{D-4}
\lim_{D'=4}(D'-4)\frac{1}{ m^{(D'-4)}} \Gamma^{(1)}[J, K_0].
\label{appD:5}
\end{eqnarray}
If we look at the defining functional equation at one loop
(\ref{fin.4.2.1p}) 
\begin{eqnarray}
&& {\cal S}_a(\G^{(1)})=\Big [
\frac{1}{2 m^{(D-4)}} \frac{\delta \G^{(0)}}{\delta \phi_a}
 \frac{\delta }{\delta K_0}
+
\frac{1}{2 m^{(D-4)}} \frac{\delta \G^{(0)}}{\delta K_0} 
\frac{\delta }{\delta \phi_a}
+\frac{1}{2} \epsilon_{abc}\phi_c \frac{\delta }{\delta \phi_b} 
\nonumber \\
&& - \partial^\mu \frac{\delta }{\delta J^\mu_a} 
   -  \epsilon_{abc} J^\mu_b \frac{\delta }{\delta J^\mu_c}
\Big ]\G^{(1)} = 0 \, 
\label{appD:6}
\end{eqnarray}
(where the rescaling (\ref{appD:1}) has been accounted for),
one sees that the same normalization and pole subtraction
as in eq. (\ref{appD:5}) should be used for amplitudes involving
only one external $\vec\phi$. In fact this can be seen on purely
dimensional grounds by counting the powers of $m$ for an arbitrary
one-loop graph after the rescaling in eq. (\ref{appD:1}).
\par\noindent
The counterterms obey the equation
\begin{eqnarray}
&& {\cal S}_a(\widehat\G^{(1)})=\Big [
\frac{1}{2 m^{(D-4)}} \frac{\delta \G^{(0)}}{\delta \phi_a}
 \frac{\delta }{\delta K_0}
+
\frac{1}{2 m^{(D-4)}} \frac{\delta \G^{(0)}}{\delta K_0} 
\frac{\delta }{\delta \phi_a}
+\frac{1}{2} \epsilon_{abc}\phi_c \frac{\delta }{\delta \phi_b} 
\nonumber \\
&& - \partial^\mu \frac{\delta }{\delta J^\mu_a} 
   -  \epsilon_{abc} J^\mu_b \frac{\delta }{\delta J^\mu_c}
\Big ]\widehat\G^{(1)} = 0 \, .
\label{appD:7}
\end{eqnarray}
%

\subsection{Two Loops}
Once the counterterms at one loop have been introduced,
the two-loop amplitudes need a further subtraction in order
to take the limit $D\to 4$.
\par
This problem can be described from different points of view.
We find it illuminating to use of the grading in the counterterms
as expressed in eq. (\ref{mod.14}) and discussed in Appendix
\ref{app:C}. Let $\Gamma^{(2,1)}$ be the vertex functional
at two loops containing the counterterms of first order 
$\widehat\Gamma^{(1)}$. Then eq. (\ref{mod.14}), after the rescaling
given in eqs. (\ref{mod.008.1p}) and (\ref{appD:1}), reads
\begin{eqnarray}
&&
\Biggl(
- \partial^\mu \frac{\delta }{\delta J^\mu_a} 
   -  \epsilon_{abc} J^\mu_b \frac{\delta }{\delta J^\mu_c}
+\frac{1}{2} \epsilon_{abc}\phi_c \frac{\delta }{\delta \phi_b} 
+\frac{1}{2m^{(D-4)}}\frac{\delta \G^{(0)}}{\delta K_0} 
\frac{\delta }{\delta\phi_a}
\nonumber \\&&
+
\frac{1}{2m^{(D-4)}}
\frac{\delta \G^{(0)}}{\delta\phi_a}
\frac{\delta }{\delta K_0} 
\Biggr)\G^{(2,1)}
\nonumber \\&&
+\frac{1}{2m^{(D-4)}}
\frac{\delta \G^{(1,0)}}{\delta K_0} 
\frac{\delta \widehat\G^{(1)}}{\delta\phi_a}
+\frac{1}{2m^{(D-4)}}
\frac{\delta\widehat \G^{(1)}}{\delta K_0} 
\frac{\delta \G^{(1,0)}}{\delta\phi_a}
  = 0 \, 
\nonumber \\&& 
\label{appD:8}
\end{eqnarray}
On the other side, $\Gamma^{(2,0)}$ obeys the functional equation
in $D$ dimensions
\begin{eqnarray}
&&
\Biggl(
- \partial^\mu \frac{\delta }{\delta J^\mu_a} 
   -  \epsilon_{abc} J^\mu_b \frac{\delta }{\delta J^\mu_c}
+\frac{1}{2} \epsilon_{abc}\phi_c \frac{\delta }{\delta \phi_b} 
+\frac{1}{2m^{(D-4)}}\frac{\delta \G^{(0)}}{\delta K_0} 
\frac{\delta }{\delta\phi_a}
\nonumber \\&& 
+
\frac{1}{2m^{(D-4)}}
\frac{\delta \G^{(0)}}{\delta\phi_a}
\frac{\delta }{\delta K_0} 
\Biggr)\G^{(2,0)}
+\frac{1}{2m^{(D-4)}}
\frac{\delta \G^{(1,0)}}{\delta K_0} 
\frac{\delta \G^{(1,0)}}{\delta\phi_a}
  = 0 \, 
\nonumber \\&& 
\label{appD:9}
\end{eqnarray}
Then $\Gamma^{(2,0)}+\Gamma^{(2,1)} $, the vertex functional at
two loop with only first order counterterms, obeys the equation
\begin{eqnarray}
&&
\Biggl(
- \partial^\mu \frac{\delta }{\delta J^\mu_a} 
   -  \epsilon_{abc} J^\mu_b \frac{\delta }{\delta J^\mu_c}
+\frac{1}{2} \epsilon_{abc}\phi_c \frac{\delta }{\delta \phi_b} 
+\frac{1}{2m^{(D-4)}}\frac{\delta \G^{(0)}}{\delta K_0} 
\frac{\delta }{\delta\phi_a}
\nonumber \\&& 
+
\frac{1}{2m^{(D-4)}}
\frac{\delta \G^{(0)}}{\delta\phi_a}
\frac{\delta }{\delta K_0} 
\Biggr)\biggl ( \G^{(2,0)}+\Gamma^{(2,1)}\biggr)
\nonumber \\&& 
+\frac{1}{2m^{(D-4)}}
\frac{\delta (\G^{(1,0)}+\widehat\G^{(1)})}{\delta K_0} 
\frac{\delta( \G^{(1,0)}+\widehat\G^{(1)})}{\delta\phi_a}
  =  
\frac{1}{2m^{(D-4)}}
\frac{\delta \widehat\G^{(1)}}{\delta K_0} 
\frac{\delta\widehat\G^{(1)}}{\delta\phi_a} \, ,
\nonumber \\&& 
\label{appD:10}
\end{eqnarray}
which agrees with eq. (\ref{mod.9}).
Two comments are in order for eq. (\ref{appD:10})
\par\noindent
i) After we normalize the amplitudes and subtract the pole parts
as described in eq. (\ref{appD:5}), the breaking term in
eq. (\ref{appD:10}) disappears.
\par\noindent
ii) Had we chosen to perform a further finite renormalization
at one loop by using the local invariant solutions of the linearized
equation in eq. (\ref{appE:4}), the pure pole structure of the
breaking term in eq. (\ref{appD:10}) would have been destroyed.
Consequently no criterion would be left at our disposal in order to choose 
the subtraction at two loops.
%

\subsection{n Loops}
A last straightforward step is necessary in order to complete
the recursive procedure of subtraction. We use again eq. 
(\ref{mod.14}) in order to find the breaking of the defining 
functional equation for $\Gamma^{(n)}$ once counterterms up
to order $n-1$ have been introduced 
\begin{eqnarray}
\Gamma^{(n)} = \sum_{k=0}^{n-1}\Gamma^{(n,k)}.
\label{appD:11}
\end{eqnarray}
The grading in the counterterms is very useful in subtracting out
the remaining singularities at $D=4$. In fact one can spot how
the n-th order counterterms enter in an essential way.
By using the identity (\ref{mod.14})
\begin{eqnarray}
&&
\Biggl(
- \partial^\mu \frac{\delta }{\delta J^\mu_a} 
   -  \epsilon_{abc} J^\mu_b \frac{\delta }{\delta J^\mu_c}
+\frac{1}{2} \epsilon_{abc}\phi_c \frac{\delta }{\delta \phi_b} 
+\frac{1}{2m^{(D-4)}}\frac{\delta \G^{(0)}}{\delta K_0} 
\frac{\delta }{\delta\phi_a}
\nonumber \\&& 
+
\frac{1}{2m^{(D-4)}}
\frac{\delta \G^{(0)}}{\delta\phi_a}
\frac{\delta }{\delta K_0} 
\Biggr)\biggl (\sum_{k=0}^{n-1}\Gamma^{(n,k)} \biggr)
\nonumber \\&& 
+\frac{1}{2m^{(D-4)}}\sum_{n'=1}^{n-1}
\frac{\delta }{\delta K_0}\biggl( \sum_{j=0}^{n'}\Gamma^{(n',j)}\biggr)
\frac{\delta }{\delta\phi_a}\biggl(\sum_{j'=0}^{n-n'}\Gamma^{(n-n',j')}\biggr)
\nonumber \\&& 
= \frac{1}{2m^{(D-4)}}\sum_{n'=1}^{n-1}
\frac{\delta }{\delta K_0}\biggl(\Gamma^{(n',n')}\biggr)
\frac{\delta}{\delta\phi_a}
\biggl(\Gamma^{(n-n',n-n')}\biggr).
\label{appD:12}
\end{eqnarray}
Since, by definition $\Gamma^{(k,k)}=\widehat\Gamma^{(k)}$, one gets
the breaking term of eq. (\ref{mod.8})
\begin{eqnarray}&&
\Delta^{(n)}
= \frac{1}{2m^{(D-4)}}\sum_{n'=1}^{n-1}
\frac{\delta }{\delta K_0}\biggl(\widehat\Gamma^{(n')}\biggr)
\frac{\delta}{\delta\phi_a}
\biggl(\widehat\Gamma^{(n-n')}\biggr). 
\label{appD:13}
\end{eqnarray}
By construction the counterterms of order $k<n$ are poles in $D-4$
multiplied by $m^{(D-4)}$, then the removal of the divergent
parts on the normalized amplitudes suppresses the breaking term
shown in eq. (\ref{appD:13}). Thus the recursive removal
of the divergences is consistent.

Let us give a closer look at the subtraction procedure.
We perform the subtraction
of the pole parts on the normalized ancestor amplitudes. The
power $\nu$ in the factor  $m^{\nu(D-4)}$ present in any $n$-th
order ancestor amplitude can be evaluated as in Appendix 
\ref{app:A}
\begin{eqnarray}&&
\nu = -I + \sum_j V^J_j + \sum_k V^{K_0}_k + \sum_l V^{\phi}_l
\nonumber \\&& 
=1-n,
\label{appD:14}
\end{eqnarray}
where $n$ is the number of loops. Thus $\Gamma^{(n)}[J,
K_0]$ behaves as  $m^{(1-n)(D-4)}$ and 
the normalized amplitudes as $m^{-n(D-4)}$. 
Consequently the removal
of the poles for the normalized amplitude  $m^{-(D-4)}
\Gamma^{(n)}[J, K_0]$ corresponds to a nontrivial choice in the
subtraction in dimensional regularization.

\end{document}